\documentstyle[11pt,IAU207_pasp,twoside,psfig]{article}
\reversemarginpar

\begin{document}
\title{Formation Scenarios for Globular Clusters and Their 
                      Host Galaxies}
\author{Stephen E. Zepf}
\affil{Dept. of Physics and Astronomy, Michigan State University, 
and Department of Astronomy, Yale University}

\begin{abstract}

	This review focuses on how galaxies and their globular
cluster systems form. I first discuss the now fairly convincing
evidence that some globular clusters form in galaxy starbursts/mergers.
One way these observations are valuable is they place important
constraints on the physics of the formation of globular
clusters. Moreover, it is natural to associate the typically
metal-rich clusters forming in mergers with the substantial
metal-rich population of globulars around ellipticals, thereby
implying an important role for galaxy mergers in
the evolution of elliptical galaxies. I also highlight some
new observational efforts aimed at constraining how and when elliptical
galaxies and their globular cluster systems formed. These include 
systematic studies of the number of globular clusters around galaxies
as a function of morphological type, studies of the kinematics
of globular cluster populations in elliptical galaxies, and
a variety of observational programs aimed at constraining the
relative ages of globular clusters within galaxies as a function
of cluster metallicity. The understanding of the formation of
globular cluster systems and their host galaxies has grown
dramatically in recent years, and the future looks equally promising.

\end{abstract}

\section{Introduction}

	The study of the formation of globular cluster systems
and their host galaxies has progressed greatly in recent years.
Two of the main subjects of this meeting - the formation of objects 
with the properties of young globular clusters in nearby galaxy 
starbursts/mergers, and the presence of distinct globular cluster 
populations around elliptical galaxies - were controversial ideas
outside of the mainstream at the last major globular cluster conference
in Santa Cruz about a decade ago. The goal of this paper is to place
these advances in the context of models of the formation of globular
cluster systems and their host galaxies, and to highlight critical
areas for future progress.
To discuss the formation of globular cluster systems and
their host galaxies is a tall order. 
To provide some focus within
this broad subject, I will orient this paper around three questions
1) How do globular clusters form (and evolve)?
2) Why do elliptical galaxies have more globular clusters than spirals 
(and how many more
do they really have)?
3) Why do many elliptical galaxy globular cluster systems have bimodal 
color distributions?

\section {Globular Cluster Formation}

	There are several reasons for starting the discussion of the 
formation of globular cluster systems and their host galaxies 
with the question of how globular clusters themselves form. 
Firstly, the physical conditions required for globular cluster formation
must have been in place during galaxy formation in order to account for
the presence of globular clusters around nearly all galaxies. If we
can determine the physical conditions that enable globular clusters to form, 
it follows that these must have been present during the formation of their 
host galaxies. Secondly, there has been dramatic recent progress observing 
the process of globular cluster formation in nearby mergers and starbursts. 
These observations directly show us at least one set of physical conditions 
that leads to the formation of star clusters with the sizes and masses 
of globular clusters.

	As evidenced in the conference poster and many talks at
this meeting, HST imaging has revealed a wealth of compact, luminous
young star clusters in starbursts and mergers. 
These star clusters
are observed to have the sizes, luminosities, and colors predicted 
for globular clusters at young ages (e.g.\ Ashman \& Zepf 1992).
Subsequent spectroscopy has confirmed their stellar nature and
ages, and in a few cases their masses inferred from the colors
and luminosities (see also many papers in these proceedings). 
Spectroscopy has also shown that the metallicities of these clusters
are very roughly solar, as expected from clusters that form out of 
enriched gas in spiral galaxy disks. 
Moreover, there are now examples of star clusters
with masses and densities like those of Galactic globular clusters
at essentially all ages. These include systems ranging from birth 
(e.g.\ NGC~4038/4039, Whitmore et al.\ 1999)
through youth at several hundred Myr (e.g.\ NGC~7252, Miller et al.\ 1997) 
to middle age at several Gyr (e.g.\ NGC~1316, Goudfrooij et al.\ 2001) to
old age (e.g.\ the Galaxy and M87, Kundu et al. 1999).

A natural interpretation of these data is 
that (metal-rich) globular clusters form in gas-rich mergers, and that 
the formation process of galaxies with a significant population of metal-rich 
globulars was similar to the nearby galaxy mergers we see today. Since 
most ellipticals have a significant metal-rich globular cluster population, 
this is essentially saying that mergers play a major role in the formation 
of ellipticals. Moreover, since roughly half of the globular clusters 
in the local universe are ``metal-rich'' ($[{\rm Fe/H}] \ga -1.0$), 
many globular clusters may have formed in a similar
fashion to that observed in nearby mergers, placing important constraints 
on GC formation models.

	These are powerful conclusions, so it is worth examining in
detail how they come about, and what are some avenues for further tests 
and exploration of these ideas. Firstly, it is important to emphasize that
the conclusion only applies directly to clusters that are fairly 
enriched, because we only observe the formation of such clusters in 
nearby starbursts and mergers. Whether the physics involved in forming
globular clusters in gas-rich mergers can be extended to low metallicity 
regimes is uncertain. This question is something Keith Ashman and I are
working on, with the main challenge being that metal-poor clusters
tend to live in lower density regions, in which it is harder to
maintain high pressures like those in dense starbursting regions.
Therefore, it is possible that low metallicity ($[{\rm Fe/H}] \la -1.0$) 
clusters have a different formation mechanism than that 
observed for high metallicity ($[{\rm Fe/H}] \ga -1.0$) clusters.
Because elliptical galaxies generally
also have a significant population of low metallicity clusters,
this also leaves open the possibility that other processes not
associated with starbursts/mergers contribute to the building up
of elliptical galaxies.

	A second question is whether all clusters with 
compact sizes and inferred masses like those of Galactic globulars 
are genuine globular clusters regardless of age. 
One concern is this regard that has
been effectively addressed is 
whether the young clusters generally lack low-mass stars
so that they will effectively disappear before reaching old age.
This arises in large part because the stellar initial mass function 
in starbursts is not well-known and also because not all of the
handful of young clusters observed with high spectral resolution 
and signal-to-noise have dynamical mass estimates consistent with
standard stellar mass functions
(Ho \& Filippenko 1996, Filippenko \& Ho 1996, Smith \& Gallagher 2001). 
However, this concern 
is soundly laid to rest by the observations of the NGC~1316 
intermediate age globular cluster system by Goudfrooij et al.\ (2001). 
These authors show convincingly through imaging and spectroscopy
that there are many bright, dense clusters in this galaxy with 
ages of about 3~Gyr. This confirms the existence of intermediate
age clusters that had been suspected based on photometry of the
cluster systems of several galaxies (e.g.\ Whitmore et al.\ 1997,
Georgakakis, Forbes, \& Brodie 2001).
The existence of 3~Gyr old star clusters is simply not 
possible if these objects only have high-mass stars.
Thus the continuous age range observed for massive, 
dense star clusters, from very young to almost a Hubble time, provides 
strong evidence that some of the dense young clusters we observe today 
will survive to ripe old age.

	The only significant observational difference between young 
and old dense cluster systems is their mass function.
Specifically, young cluster systems appear to have a mass function  
that is a power-law with increasing numbers down to the limit of 
the observations of a few $10^4 M_{\sun}$, while old systems have a
mass function that is similar to that of the young systems for 
masses greater than 1-2~$10^5 M_{\sun}$, but below this ``turnover'' 
mass the number of objects in old systems follows a very much
shallower slope, so that it has many fewer low-mass objects.
This difference in mass function has now been established for a 
number of both young and old systems
(e.g.\ Fall \& Zhang 1999, Zepf et al.\ 1999,
Carlson et al.\ 1998, Miller et al.\ 1997). The question is whether 
this difference reflects a fundamental physical difference between 
younger and older globular cluster systems, or whether this is a 
result of the dynamical evolution of cluster systems. 

	Dynamical evolution of globular clusters is a consequence 
of basic physics. It has also long been realized that dynamical effects 
will tend to destroy lower mass clusters preferentially (e.g.\ Fall \& Rees 
1977, Spitzer 1987). Therefore, there is little question that in a
qualitative sense, basic gravitational physics will tend to turn an 
initially power-law cluster mass function like that seen in young
systems into something resembling the log-normal shape typical of
old globular cluster systems. However, whether this actually 
works quantitatively remains a critical unanswered question. 
On the theoretical side, there has been much recent work on the
subject (see talks at this meeting by Fall and Vesperini), but
there is not yet a full consensus on whether dynamical evolution
can produce a luminosity function as uniform as observed between
galaxies and within galaxies from an initial mass function that is
a power-law rising steeply to low masses. 
Observationally, perhaps the most critical question is
whether intermediate cases can be found between the power-laws
with no turnover down to several $10^4 M_{\sun}$
in young systems and the log-normal distributions with a ``turnover''
around 1-2~$10^5 M_{\sun}$ in old systems. As discussed in Zepf et al.\ (1999) 
it is not necessarily surprising that no turnover has been found in
current studies of young systems, and it will be hard to push these 
to lower masses because of contamination with individual stars
(cf.\ Whitmore et al.\ 1999). The best bet to make this test seems 
to be to push observations of intermediate age systems of several Gyr 
old to very faint levels.

	The fact that globular cluster formation is directly observed
is also a great advance for the study of how globular clusters form.
Prior to these observations, there were few direct constraints on
the physical conditions that lead to globular cluster formation.
The observations of globular cluster formation in nearby starbursts 
and mergers have changed this situation dramatically.
One attempt to take advantage of these new observations to inform
models of globular cluster formation is a paper Keith Ashman and I 
have recently written (Ashman \& Zepf 2001).
A key point of our paper is that the high pressure observed
in starbursts implies that any bound clouds that form in the ISM of the
starburst will be much more compact than typical Galactic molecular
clouds, and that for typical starburst pressures, clouds with globular
cluster-like masses will have globular cluster-like radii. We also
show that the recent observation of Zepf et al.\ (1999) that
the young star clusters have a weak or absent mass-radius relation places
strong constraints on formation models, since nearly all models 
start from clouds with a mass-radius relation. Specifically, we find 
that if the star formation efficiency in the progenitor cloud 
scales with the binding energy of the cloud,
star clusters without a mass-radius relation may be formed from
clouds with the virial mass-radius scaling observed for Galactic 
molecular clouds. 

\section{Why do elliptical galaxies have more GCs than spirals?} 

	The origin and implications of the greater specific
frequency (number normalized by galaxy luminosity or mass) of
globular clusters around elliptical galaxies compared to spiral
galaxies is a long-standing question. This difference was first 
used to argue that ellipticals are not formed by the mergers of 
spirals (e.g.\ van den Bergh 1990, Harris 1991), since the 
simple combination of spirals would result in ellipticals with 
the same normalized number of globular clusters as spirals. 
Schweizer (1987) and Ashman \& Zepf (1992) argued that globular 
clusters could form in gas-rich mergers,
thereby resulting in more globular clusters around ellipticals
relative to spirals if ellipticals form by merging spirals.
Moreover, Ashman \& Zepf (1992) predicted that this would 
lead to bimodal metallicity distributions for globular cluster systems
around elliptical galaxies. Specifically, elliptical galaxies
formed by mergers would have a metal-poor population from the halos 
of the progenitor spiral galaxies, and a metal-rich population formed 
during the merger from the enriched gas in the disks of the spiral 
galaxies.

	However, before venturing too far into the discussion of
why elliptical galaxies have more globulars than spiral galaxies,
it is useful to consider the actual observational constraints
on the number of globular clusters around different galaxies.
In Figure 1, I plot an updated version of the number of globular
clusters normalized by the stellar mass of the galaxy against
galaxy mass (cf. Ashman \& Zepf 1998). The overall trend that 
elliptical galaxies, represented by open symbols, appear to lie 
above spiral galaxies, represented by closed symbols, seems
clear from this figure. Thus, there is little question that
per unit stellar mass, elliptical galaxies have more globular
clusters than spiral galaxies. However, there is also tremendous
scatter in this diagram.

\begin{figure}
\vspace{-0.8in}
\psfig{figure=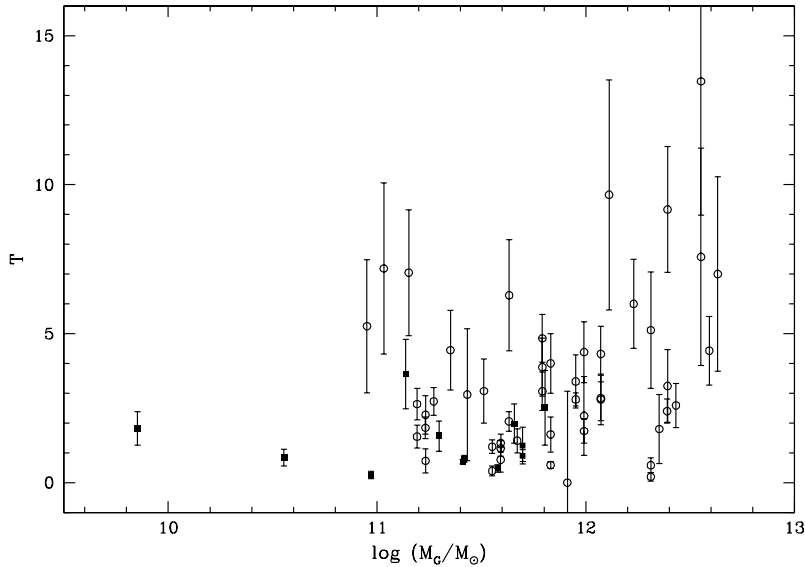,width=13cm}
\vskip-1.39in
\caption{The mass-normalized specific frequency $T$
plotted against estimated galaxy mass. $T$ is defined as 
$T \equiv N_{GC}/(M_{G}/10^9 M_{\sun})$.
Early-type galaxies are
plotted as open circles and spiral galaxies as filled squares. The
error bars are only estimates of the statistical uncertainties and
do not account for a number of possible systematic concerns. Most
of the data are from Ashman \& Zepf (1998) where references are listed, 
with several updates from the literature that have generally decreased 
$T$ for some of the massive ellipticals.
\vspace{-0.3truein}  
}
\end{figure}

	What limits the comparison of ellipticals and spirals in
this diagram? Firstly, there are very few well-studied spirals.
The Milky Way and M31 dominate the comparison, along with
M104, if it is counted as a spiral despite its very large bulge-to-disk 
ratio. This problem is being addressed in Katherine Rhode's PhD thesis,
which includes high quality WIYN data for nine spiral galaxies.
However, for now, the elliptical to spiral comparison suffers from
having only a handful of well-studied spirals. A second major problem
is the large uncertainties in the elliptical galaxy measurements
of the specific frequency of globulars. This can be seen by the sizeable
error bars in Figure 1. Moreover, these probably underestimate
the true errors because they typically only account for statistical
uncertainties and not systematic concerns such as the extrapolations
to large radii often necessary to estimate the total number of 
globulars. Mosaic Camera data for the well-studied
Virgo elliptical NGC~4472 suggest that its specific frequency is
about $30\%$ less than previously assumed (Rhode \& Zepf 2001). 
Similar results may now be coming out for M87 (Forte at this meeting).
The obvious way to address this problem is with a systematic
multi-color survey of early-type galaxies with the new generation 
of CCD Mosaic imagers.

	Despite the uncertainties, there are probably still some
trends that show up in the data presented in Figure 1. One of these
is that the specific frequency of globular clusters appears to
increase slightly with increasing elliptical galaxy luminosity/mass
(e.g.\ Djorgovski \& Santiago 1992; Zepf, Geisler, \& Ashman 1994,
Kissler-Patig 1997).
There are a number of plausible explanations for this effect.
One possibility is dynamical evolution, because more luminous ellipticals
have shallower density gradients, and thus destroy a smaller fraction of
their clusters (e.g.\ Murali \& Weinberg 1997).
Therefore, dynamical evolution will naturally produce at least some
of the observed trend of $T$ with elliptical galaxy mass/luminosity. 
Another possibility is based on evidence that more massive ellipticals 
tend to have more gas mass. Therefore, the systematic trend of $T$ for
massive ellipticals might be accounted for by adopting a constant
number of globular clusters per baryon mass, rather than per stellar 
mass (e.g.\ Blakeslee, Tonry, \& Metzger 1997; McLaughlin 1999).
Although the hope that the globular cluster frequency per ``baryon mass'' 
might be constant within a galaxy has also been ruled out recently 
in the best studied case of NGC~4472 (Rhode \& Zepf 2001), there
is a good correlation for the most massive ellipticals between
high specific frequency and large gas mass.

	However, the most striking result of Figure 1, the systematic
difference between the mass-normalized frequency of globular clusters
around ellipticals relative to spirals, is hard to account for with 
either of these effects. The appeal to the hot gas as ``missing stars'' 
only works for massive ellipticals, as typical ellipticals around 
$L_{\star}$ do not have much mass in hot gas and may have a significant
contribution to their X-ray luminosity from stellar sources 
(e.g.\ Sarazin, Irwin, \& Bregman 2000).
Therefore, the difference in the mass-normalized globular cluster frequency 
($T$) between ellipticals and spirals of the same stellar mass is not 
solved by including mass in hot gas.
Dynamical evolution is more physically plausible, since the disks
and compact bulges of spiral galaxies can accelerate the destruction
of clusters (e.g.\ Gnedin \& Ostriker 1997, Vesperini 1998), potentially 
resulting in fewer clusters around spirals compared to ellipticals.
However, if the dynamical effects are strong enough to account for 
differences in specific frequency as large as those observed between 
ellipticals and spirals, they tend to produce significant differences 
between the mass functions of the different systems (e.g.\ Vesperini 2001). 
The strong similarity of the globular cluster luminosity functions of 
ellipticals and spirals (e.g.\ Whitmore 1997) therefore argues
against dynamical evolution as the primary driver of the $T$ difference
between spirals and ellipticals.
Finally, it is possible to normalize the globular cluster numbers
to bulge or halo luminosity (e.g.\ Harris 1991). In this case, the
specific frequency of spiral bulges and/or halos can be similar to 
that of ellipticals (e.g.\ Forbes 2001), depending in large part 
on how the difficult task of determining bulge and/or halo luminosities 
is done for late-type galaxies. 
However, this approach does not address the critical question 
why disks like those of the Milky Way and M31 are unfavorable 
for globular cluster formation while the events that created
the main bodies of elliptical galaxies were efficient at globular
cluster formation, and what this means for how these galaxies formed.

	The most appealing answer would seem to take a lesson from
observations that quiescently star-forming disks are
unfavorable for globular cluster formation, but actively starbursting
and merging systems form globular clusters efficiently. Thus, if
elliptical galaxies are those objects that had major mergers in their
past, while spirals avoided such catastrophic events, the difference
in their specific frequencies might be accounted for simply through the
fact elliptical galaxies formed more of their stars in a mode favorable 
for globular cluster formation (Ashman \& Zepf 1992, Zepf \& Ashman 1993).
Of course, such a proposal needs to be tested and compared to 
alternative pictures.
Several efforts along these lines are discussed in the next section.

\section{Why are the globular cluster systems of elliptical galaxies bimodal?}

	Ashman \& Zepf (1992) suggested that elliptical galaxies
formed by the mergers of disk galaxies would have two populations
of globular clusters - a metal-poor population from the halos 
of the progenitor spiral galaxies, and a metal-rich population formed 
during the merger from the enriched gas in the disks of the spiral
galaxies. This prediction is much different than the expectation
of classical monolithic collapse models for elliptical galaxy formation
in which a single metallicity peak is expected (e.g.\ Arimoto \& Yoshii 1987).
Therefore, when bimodality was first discovered in the color distribution
of elliptical galaxy globular cluster systems (Zepf \& Ashman 1993,
Ostrov, Geisler, \& Forte 1993), it was a striking success for a
{\it prediction} of a simple merger model. 

	Although the prediction of the simple merger model leads
to bimodality, it is natural to ask what happens when one considers
a more complex merger history involving several different major
mergers over time, along with many lesser accretion events.
While a complete calculation along these lines would require
a detailed understanding of a number of challenging subjects such
as feedback and chemical evolution, some progress can be made by
treating the Milky Way as a typical disk galaxy, and considering
what we know about its metallicity as a function of time.
The critical aspect of the Milky Way in this regard is that the
disk was enriched to near its current metallicity very early in 
its history (Ng \& Bertelli 1998; Carraro, Ng, Portinari \& 1998).
If this is typical of disk galaxies, then it is natural that most
of the globulars formed in mergers will have fairly high 
metallicity. It also follows from this argument that mergers
of enriched, gas-rich disks will form elliptical galaxies with
mostly metal-rich stars. This is consistent with observations
that elliptical galaxies appear to have a G-dwarf problem
(i.e.\ an absence of low metallicity  stars compared to closed box models) 
similar to that in spirals (e.g.\ Harris \& Harris 2000, Marleau et al.\ 2000, 
Worthey, Dorman, \& Jones 1996). 

	Thus it seems plausible that even complex merger histories
will produce globular cluster systems of many elliptical galaxies 
with generally bimodal color distributions. This arises
because disks like that of the Milky Way were enriched to near
solar metallicity very early, and thus mergers of these disks will
produce clusters that are predominantly metal-rich, while the halos
of the progenitor spirals have clusters that are predominantly
metal-poor.
Of course, globular clusters formed out of enriched gas in early
disks will be unlikely to all have exactly the same metallicity.
There will likely be radial gradients in the progenitor disks,
and there will probably be a dependence of the metallicity
of the disk on the galaxy luminosity/mass.
In a merger-induced starburst, there
will likely be additional enrichment, the amount of which will
probably be tied to the depth of the potential well.
Moreover, a few gas-rich mergers will have happened recently
enough that the metal-rich clusters will be too young to be 
clearly redder than the old metal-poor clusters.
However, observing these fine distinctions in typical color
distributions will be very difficult because of photometric errors and
uncertainties in the conversion of broad-band colors to metallicity.
The result is that the roughly solar metallicity of the Milky Way disk 
over nearly all of the Hubble time suggests that most of the gas-rich
merger events that build up elliptical galaxies and possibly other
large bulge systems will produce globular clusters with metallicities
somewhere near solar, and these may often result in bimodal color 
distributions for globular cluster systems.

	Another natural question is the origin of the metal-poor
cluster populations around early-type galaxies. In the simplest merger 
picture, the metal-poor clusters come from the halos of the progenitor 
spirals. However, it has long been recognized that this model appears 
to be inadequate to account for the full range of properties of the 
metal-poor cluster populations around ellipticals, if the halo populations 
of the Milky Way and M31 are typical of such progenitor spirals
(e.g. Zepf, Ashman, \& Geisler 1995, Forbes et al.\ 1997, 
Kissler-Patig et al.\ 1998, C\^ot\'e, Marzke, \& West 1998, 
Ashman \& Zepf 1998).
In particular, a few ellipticals appear to be ``missing'' a significant 
metal-poor population (e.g.\ Woodworth \& Harris 2000, Zepf et al.\ 1995), 
while many very massive ellipticals appear
to have a high specific frequency of metal-poor clusters.
One of the weaknesses in all of these studies are the large and
potentially systematic uncertainties in specific frequencies
discussed in the previous section. For example, NGC~4472 has
often been taken as a fiducial example of a massive elliptical
with a high specific frequency, but better data from Mosaic CCD
imaging suggests that the specific frequency of this galaxy is 
lower than previous estimates (Rhode \& Zepf 2001). 
Moreover, all of these analyses rely on the assumption that the 
Milky Way and M31 halos are typical of spiral galaxies.
However, even with these serious systematic concerns,
there are almost certainly some very massive ellipticals with
higher mass-normalized specific frequencies for metal-poor clusters 
than can easily be achieved by combining typical spiral galaxies.

	There are several possibilities for accounting for the
large number of metal-poor globulars around some massive ellipticals.
One is to consider that even the halos of the Milky Way and M31
appear to have been at least partially the result of accretion
and merging over time. These processes would also be expected
to occur both in the halos of the progenitor spirals and in the
elliptical galaxy after it formed from major mergers. If for
some reason,
these elliptical galaxies
accreted more low-mass galaxies, or the accreted galaxies had
a higher specific frequency of metal-poor globulars, then
it would be possible to make an elliptical with a larger 
mass-normalized specific frequency of metal-poor globulars. 
This process was modeled in detail by C\^ot\'e et al.\ (1998), who
found they could match the estimate of the metal-poor globular 
cluster population of NGC~4472 at that time, with a steep galaxy luminosity 
function ($\alpha \simeq -1.8$) and a fairly high specific frequency 
of metal-poor globulars for dwarf galaxies.
An alternative is to revisit the idea that the gas mass around
the most massive ellipticals was available to form metal-poor globulars,
but not stars (e.g.\ Blakeslee et al.\ 1997,
Harris, Harris, \& McLaughlin 1998; McLaughlin 1999)
These two ideas are not necessarily separate, since they both
can be more or less accommodated in a model with a population
of dwarf galaxies around these massive ellipticals that produces
some globulars and not many stars in a burst, and then disperses,
and is unable to cool again.

	Perhaps the most interesting question is what observations
can be made to shed further light on how elliptical galaxies and
their globular cluster systems formed. Here I will highlight two
such possibilities. One area that can shed light on the physical
mechanisms involved in the formation of elliptical galaxies and
their cluster systems is the study of the kinematics of the globular
cluster systems. Because this is the subject of other recent reviews
(e.g.\ Bridges 2001, Zepf 2001), I will only highlight one aspect
of this work. A key question that the kinematics can address
is the angular momentum of globular cluster populations. If galaxies
generally form by dissipational collapse, they will spin-up
as the result of the collapse and conservation of angular momentum
(e.g.\ Fall \& Efstathiou 1980). In contrast, mergers provide a way 
to transport angular momentum outwards (e.g.\ Barnes \& Hernquist 1992).
Therefore, it is potentially very interesting that of the three
elliptical galaxies for which fairly detailed kinematics of the
globular cluster systems exist (NGC~5128, Hui et al.\ 1995; M87, 
Cohen 2000, C\^ot\'e et al.\ 2001; and NGC~4472, Zepf et al.\ 2000),
the rotation is significantly smaller than the dispersion for
two of these (M87 and NGC~4472). In both cases, rotation appears
to become more important at larger radii and/or for the less
centrally concentrated metal-poor cluster populations, also
suggesting angular momentum transport.

	Another critical observation are the relative ages of the
metal-rich and metal-poor clusters. In general, merger models predict
that the typical ages of the metal-rich clusters will be younger than
the typical ages of the metal-poor clusters. Conversely, models
that start with a monolithic collapse at the center, and accrete the
metal-poor population afterwards tend to predict older ages for the
metal-rich clusters compared to the metal-poor clusters. 
The challenge is that determining ages, even relatively, is difficult.
Even for the Galaxy and M31 the situation is not completely
clear, although there is tentative evidence that at least
some of the metal-rich globulars in the Milky Way (Rosenberg et al.\ 1999)
and M31 (Barmby, Huchra, \& Brodie 2001) are somewhat
younger than the typical metal-poor globulars in these galaxies. 
Unsurprisingly, the situation is more difficult for globulars
outside of the Local Group. Here I will review several of the
techniques being applied, although they are far from definitive
as yet. One common approach is to use the Balmer lines
as age indicators. However, when applied to 47~Tuc and other metal-rich
Galactic globulars, this gives ages
of $\sim 20$ Gyr (Gibson et al.\ 1999, Vazdekis et al.\ 2001), 
so there are clearly some unresolved systematic questions 
(cf. Cohen, Blakeslee, \& Ryzhov 1998, Beasley et al.\ 2000).
In general the studies above 
find that the ages may be similar
for cluster populations of different metallicities and that
no cluster population is very young. 

	Another approach to determining the age differences between
metal-rich and metal-poor populations is to compare their luminosity
functions. Given the assumption the underlying mass functions are
the same, the difference in the luminosity function can be turned
into a difference in mass-to-light ratio, which can be combined
with the observed color difference to estimate the relative
age and metallicity of each population. When applied to M87,
this technique appears to suggest the metal-rich population
is somewhat younger than the metal-poor population (Kundu et al.\ 1999),
and when applied to NGC~4472, the data tend to suggest fairly
equal ages (e.g.\ Puzia et al. 1999). As discussed in these papers,
the errror bars on these techniques are significant, and the
results are very dependent on the stellar populations models
applied and the assumption of equivalent mass functions for
different metallicity populations.

	An approach along somewhat similar lines is to use a 
wide enough spread of broad-band colors so that the age-metallicity
degeneracy is broken to some extent. Puzia et al.\ (2001) and
Kissler-Patig et al.\ (2001) have attempted this
using $VIK$ colors with some success.
An independent
approach is to use far-UV colors as a probe of the color of the
horizontal branch, which also has an age dependence. Two HST
studies along these lines (PIs Zepf and O'Connell) are being
carried out now, and the results should be forthcoming next year.
Overall, there are suggestions of modestly younger, metal-rich
populations in elliptical galaxies, but it is still early
days for this work.










\acknowledgments
	
	This work is due in large part to my many excellent
collaborators. I especially acknowledge my long-time collaboration
with Keith Ashman. Some of the work presented here has been supported 
by NASA Long-Term Space Astrophysics grant NAG5-9651 and NASA
Astrophysics Theory Program grant NAG5-9168, and my travel
to this meeting was partially supported by an AAS travel grant.


\begin{references}

\reference
Arimoto, N., \& Yoshii, Y. 1987, A\&A, 173, 23

\reference
Ashman, K.M., \& Zepf, S.E. 1992, ApJ, 384, 50

\reference
Ashman, K.M., \& Zepf, S.E. 1998, Globular Cluster Systems
(Cambridge: Cambridge University Press)

\reference
Ashman, K.M., \& Zepf, S.E. 2001, AJ, in press (astro-ph/0107146)

\reference
Barmby, P., Huchra, J.P., \& Brodie, J.P. 2001, AJ, 121, 1482

\reference
Barnes, J. E., \& Hernquist, L. 1992, ARA\&A, 30, 705 

\reference
Beasley, M.A., et al.\ 2000, MNRAS, 318, 1249

\reference
Blakeslee, J. P., Tonry, J. L., \& Metzger, M. R. 1997, AJ, 114, 482 

\reference
Bridges, T.J. 2001, these proceedings

\reference
Carlson, M.N. et al.\ 1998, AJ, 115, 1778

\reference
Carraro, G., Ng, Y.K., \& Portinari, L. 1998, MNRAS, 296, 1045

\reference
Cohen, J.G. 2000, AJ, 119, 162

\reference
Cohen, J.G., Blakeslee, J. P., \& Ryzhov, A. 1998, ApJ, 496, 808 


\reference
C\^ot\'e, P., Marzke, R.O., \& West, M.J. 1998, ApJ, 501, 554

\reference
C\^ot\'e, P., et al.\ 2001, ApJ, in press

\reference
Djorgovski, S.G., \& Santiago, B.X. 1992, ApJ, 391, L85

\reference

\reference
Fall, S.M., \& Efstathiou, G. 1980, MNRAS, 193, 189

\reference
Forbes, D.A., 2001, these proceedings

\reference
Forbes, D.A., Brodie, J.P., \& Grillmair, C.J. 1997, AJ, 113, 1652

\reference
Georgakakis, A.E., Forbes, D.A., \& Brodie, J.P. 2001, MNRAS, 324, 785

\reference
Gibson, B., et al.\ 1999, AJ, 118, 1268

\reference
Gnedin, O.Y., \& Ostriker, J.P. 1997, ApJ, 474, 223

\reference
Goudfrooij, P., Mack, J., Kissler-Patig, M., Meylan, G., \& Minniti, D.
2001, MNRAS, 322, 643

\reference
Harris, G.L.H., \& Harris, W.E. 2000, AJ, 120, 2423

\reference
Harris, W.E. 1991, ARAA, 29, 543

\reference
Harris, W.E., Harris, G.L.H., \& McLaughlin, D.E. 1998, AJ, 115, 1801

\reference
Kissler-Patig, M. 1997, A\&A, 319, 83

\reference
Kissler-Patig, M., Brodie, J.P., \& Minniti, D. 2001, A\&A, in preparation




\reference
Marleau, F. R., Graham, J. R., Liu, M. C., \& Charlot, S. 2000, AJ, 120, 1779

\reference
McLaughlin, D.E. 1999, AJ, 117, 2398

\reference
Miller, B.W., Whitmore, B.C., Schweizer, F., \& Fall, S.M. 1997, AJ, 114, 2381

\reference
Murali, C., \& Weinberg M. 1997, MNRAS, 288, 767

\reference
Ng, Y.K. , \& Bertelli, G. 1998, A\&A, 329, 943

\reference
Ostrov, P., Geisler, D., \& Forte, J. C. 1993, 105, 1762

\reference
Puzia, T.H., Kissler-Patig, M., Brodie, J.P., \& Huchra, J.P. 1999, AJ, 118, 
2734

\reference
Puzia, T.H., Zepf, S.E., Kissler-Patig, M., Hilker, M., Minniti, D., \&
Goudfrooij, P. 2001, A\&A, in preparation 

\reference
Rhode, K.L., \& Zepf, S.E. 2001, AJ, 121, 210

\reference
Rosenberg, A., Saviane, I., Piotto, G., \& Aparicio, A. 1999, AJ, 118, 2306 

\reference
Sarazin, C. L., Irwin, J. A., \& Bregman, J.N. 2000, ApJ, 544, 101

\reference
Smith, L.J., \& Gallagher, J.S. 2001, MNRAS, in press (astro-ph/0104429)

\reference
Spitzer, L. 1987, Dynamical Evolution of Globular Clusters 
(Princeton: Princeton University Press)

\reference
van den Bergh, S. 1990, in Dynamics and Interactions of Galaxies,
ed. R Wielen (Berlin: Spinger), 492

\reference
Vazdekis, A., Salaris, M., Arimoto, N., \& Rose, J.A. 2001, AJ, 549, 274

\reference
Vesperini, E. 1998, MNRAS, 299, 1019

\reference
Vesperini, E. 2001, MNRAS, 322, 247

\reference
Whitmore, B.C. 1997, The Extragalactic Distance Scale, ed. M. Livio
(Cambridge: Cambridge University Press), 254

\reference
Whitmore, B.C., Miller, B.W., Schweizer, F., \& Fall, S.M. 1997, AJ, 114, 1797

\reference
Whitmore, B.C., et al.\ 1999, AJ, 118, 1551

\reference
Woodworth, S.C., \& Harris, W.E. 2000, AJ, 119, 2699

\reference
Worthey, G., Dorman, B., \& Jones, L. A. 1996, AJ, 112, 948


\reference
Zepf, S.E. 2001, in The Shapes of Galaxies and Their Halos, 
in press

\reference
Zepf, S.E., \& Ashman, K.M. 1993 MNRAS, 264, 611

\reference
Zepf, S.E., Ashman, K.M., \& Geisler, D. 1995, ApJ,  443, 570

\reference
Zepf, S.E., Geisler, D., \& Ashman, K.M. 1994, ApJ, 435, L117

\reference
Zepf, S.E., et al.\ 1999, AJ, 118, 752

\reference
Zepf, S.E., et al.\ 2000, AJ, 120, 2928

\reference
Zhang, Q., \& Fall, S.M. 1999, ApJ, 527, 81

\end{references}
\end{document}